# Recognition of COVID-19 Disease Utilizing X-Ray Imaging of the Chest using CNN

Md Gulzar Hussain[1], Ye Shiren[2]
School of Computer and Artificial Intelligence, Changzhou University Changzhou,
Jiangsu, China.
Email: [1]gulzar.ace@gmail.com, [2]yes@cczu.edu.cn

*Abstract*—Since this COVID-19 pandemic thrives, the utilization of X-Ray imaging (CXR) of the Chest as a complementary screening technique to RT-PCR testing grows to its clinical use for respiratory complaints. Many new Deep Learning approaches have been established for this reason. The goal of this research is to evaluate the convolutional neural networks (CNNs) for COVID-19 classification using chest X-rays. The performance of CNN with one, three, and four convolution layers is evaluated in this research. A dataset of 13,808 CXR photographs are used in this research. When evaluated on X-ray images with three splits of the dataset, our preliminary experimental results show that the CNN model with three convolution layers can reliably detect 96 percent accuracy (precesion being 96 percent). Which indicates the promise of our suggested model for reliable COVID19 screening.

*Keywords—COVID-19, X-Ray, Images Classification, CNN.*

## I. INTRODUCTION

COVID-19 is an aggressive sickness aggravated by the coronavirus type namedsevere acute respiratory syndrome coronavirus 2 (SARS-CoV-2). The health and economic consequences of the coronavirus disease 2019 (COVID-19) pandemic seem to have been unprecedented. It has infected approximately 175.69 million individuals worldwide, resulting in approximately 38.04 million fatalities [1]. This pandemic has had a severe effect on the worldwide healthcare system in this perspective. Despite all the efforts, strong public transmission is firmly established in many countries and populations [2]. The prompt and precise detection of infected individuals is crucial in the fight over COVID-19. Reverse transcriptionpolymerase chain reaction (RT-PCR) is the most commonly used method for diagnosing COVID-19, but it has low accuracy, latency, and sensitivity [3]. Chest X-ray (CXR) radiographic screenings are a supplementary screening approach to RT-PCR that is gaining popularity and utilization in clinical institutes around the world. Others have validated the use of radiography as a source of information in enabling the rapid identification of COVID-19 [4]. Identifying COVID-19 with strong precision using X-ray of chest is difficult, not only because of the ribs underlying soft tissue and poor resolution but also for the very little availability of a huge number of tagged data. It is especially true for deep learning-based approaches, which are extremely hungry for data. Motivated by the immediate need to create solutions to assist in the struggle against the COVID-19 epidemic, and motivated by the scientific community's open source and open access activities, the performance of CNN with 1, 3, and 4 convolution layers are evaluated on X-ray images of chest. The goal of this research was to look at the possibility of parameter tweaks in CNNs with 1, 3, and 4 convolution layers. If these fine-tuned networks can reach desirable performance in the recognition of COVID-19 X-ray photographs of chest by configuring them in such a way that they do the task well, the discoveries will make a substantial contribution to coronavirus epidemic relief. The discovery of the relatively simple yet powerful performance of basic fine-tuned CNNs that can yield superior accuracy in identifying COVID-19 X-ray imagery of chest with less training effort than other established deep-learning models is the research'ssignificant contribution.

## II. RELATED WORKS

Deep learning has brought a fresh approach to overcoming pandemic difficulties since its inception. Using 2D convolutional neural networks (CNNs) as a screening tool for early disease diagnosis has been a fascinating field of research. Authors of [3] tried to experiment five image enhancement techniques to detect COVID-19. Their data-set contained 18479 CXR images among them 8851 is of normal, 6012 is of non-COVID and 3616 if of COVID positive. Six different CNNs are used to investigate the performances and their Unet model gives 98.6% accuracy which outperformed others for lung segmentation. Authors of [5] proposed a sturdy technique to recognize COVID-19 automatically using digital X-ray images of chest. They collected various COVID19 datasets from public datasets and merge them to use in their research. 1579 normal, 423 COVID-19, and 1485 viral pneumonia X-ray images were available in their dataset. Different Deep CNN models are used in their binary classifier problem to analyse the performance. They found around 99 percent accuracy in their research. It is aimed to develop an alternative model emerged on capsule network to classify covid-19 x-ray images and used two publicly available dataset. Their model performed better than CNN based models with accuracy 95.7% and when pretrained 98.3% [6]. Authors introduced COVID-Net CXR-S to predict airspace severity of a covid-19 patient emerged on their chest x-ray images. around 16000 images were in their dataset. CheXNet, ResNet50 are used to compare with their model COVID-Net CXR-
S. they found that their proposed mode COVID-Net CXR-
S performed with accuracy of 92.66% [7]. Authors of [8] introduced COVID-Net CXR-2 to predict airspace severity of a covid-19 patiant emerged on their chest x-ray images and around 19203 images were in their dataset. COVID-Ne, ResNet-50 are used to compare with their model COVID-Net CXR-2. They found that their proposed mode COVID-Net CXR-2 performed with accuracy of 96.3%. COVID-Net CT-S to assess the severity of lung disease due to COVID19 using CT images in this work [9].China National Center for Bioinformation (CNCB) dataset is used for their research. 2D CT-S50, COVID-Net CT-S152, COVID-Net CT-S100 and COVID-Net CT-S50 used for performance testing. COVIDNet CT-S152 outperformed others with accuricy of 78.5%. Authors proposed a light weighted shallow CNN architecture to detect COVID-19 positive cases using CXR images [10]. 321 Covid positive and 5856 non covid images were used which are available publicly. Their proposed architecture performed better by getting 99.69% accuracy. Authors intended to detect corona virus 2019 infection based on chest radiography images. collected 1200 chest radiography images from two publicly available datasets. ShuffleNet and SqueezeNet based architecture and multiclass support vector machine classifier is proposed in their research. 96.7% accuracy for COVID19 was achieved for their proposed system. Authors [11] tried to detect COVID-19, non-Covid and Healthy cases from chest x-ray images. They used data from RSNA Pneumonia detection and COVIDx datasets. They used AutoEncoder to extract infromation from images and deep CNN to classify them. They found 93.5% accuracy for their proposed system. Authors of [12] tried to solve the data imbalance problem of x-ray image classification problem. Two different benchmark datasets were used to test their proposed architechture. Their CNN is designed to solve gradiant decent problem and in different layer features are combined dinamically to improve the classification performance. Their proposed architecture gained 99.6% accuracy. Paper [13] introduced a new CNN model for detecting covid-19 in x-ray photographs. Their dataset consist of 13975 chest x-ray photographs which was gathered from five different repositories. VGG-19, ResNet-50 and COVID-Net models were used in their research. Covid-net performed better than others with accuracy of 93.3%. In paper [14] authors used 1531 images in which 1078 is confirmed as covid-19 and others are non covid x-ray photographs. They suggested a modified deep liearning model for their research. Their models is based on annomaly detection. They found their model gives 96% accuracy for covid-19 images and 70.65% for non covid image datasets. Paper [15] used 80 normal, 20 sars and 105 covid-19 cxr images for their research. They used decomposed, transfered and composed deep CNN to classify COVID-19 cxr chest images and found 93.1% accuracy in their research. Authors of [16] classifed normal and covid-19 x-ray images using deep CNN pre-trained models. They used 180 covid-19 and 200 normal x-ray images for their research. Deep CNN pretrained models ResNet50, ResNet18, ResNet101, VGG19, and VGG16 are utilized in their research. They found 92.6% accuracy for ResNet50 model. Authors tried to compare performance of three pretrained models of CNN on Covid -19 x-ray datasets in [17] and used 3 public x-ray datasets. Three pretrained CNN models AlexNet, GoogleNet, and SqueezeNet are used in their research. For different datasets different models performed well which is around 99 percent. Authors of [18] found the relationship between ROIs in CXR images, used 3 datasets, and proposed VGG-16 model to classify Covid-19. In paper [19], the combination of ASSOA and MLP algorithm achieved 99.0% accuracy to classify X-ray COVID-19 found on GitHub.

### III. PROPOSED METHODOLOGY

A convolutional neural network with different layers is applied to recognize COVID-19 using chest X-ray photographs in this study. More specifically, a 2D convolutional neural network with 1, 3, and 4 layers implemented to use in this work. The suggested research work's system flow diagram is shown in Figure 1. The subsequent subsections go over the specifics of data collection network configuration, and performance validation.

*A. Data Collection*

Even though there are a massive proportion of COVID19 individuals suffering, the amount of publicly accessible chest X-ray photographs on the internet is small and scattered. This study made use of a publicly available dataset of chest X-ray photographs of COVID-19 positive patients with Normal as well as Viral Pneumonia. [5] [3]. The COVID-19 radiography data-base contains chest X-rays images of 3616 COVID-19 affirmative cases, as well as 10,192 Normal, 1345 Viral Pneumonia, and 6012 Lung Opacity (Non-COVID lung infection) photographs. Among them, chest X-rays images of 3616 COVID-19 affirmative cases and 10,192 Normal, a total of 13,808 images of chest X-rays are utilized in this research to recognize COVID-19. Though the images were 299 x 299 pixels, the size is reduced to 30 x 30 pixels for this research work. It is done to fit them to CNN.

*B. Preprocessing and Data Augmentation*

The data-sets were pre-processed to alter the size of the X-Ray photographs to fit the CNN model's input image size specifications, which are 30 x 30 pixels for this network. After resizing the data, it is preprocessed by rearrangement of it into the structure expected by the network model and mounting it such that all magnitudes are in the [0, 1] limit. Previously, for example, training data were retained in an array of size (13808, 28, 28) of type uint8 with values ranging from 0 to 255. It is converted into a float32 array of size (13808, 28 * 28) with values ranging from 0 to 1.

*1) Data Augmentation:* Data augmentation can increase the classification performance of deep learning models by augmenting available data. Data augmentation can significantly improve the amount of data provided for training models. When the data-set is unbalanced, image augmentation is critical. Data augmentation creates more training data from previous training samples by augmenting them with a series of random transformations that produce believable-looking images. This allows the model to be exposed to more dimensions of the data and categorize more effectively. In this research, images of normal cases are 10,192 which is more than thrice than the COVID-19 images. So, it is critical to augment the images to balance the data-set. In this research, an image rotating, shifting, shearing and zooming based augmentation approach was used to create COVID-19 training images before applying them to CNN models for training.

### C. Convolutional Neural Networks

A convolutional neural network (CNN) is a form of artificial neural network that is developed particularly for processing pixel data in image detection and processing [20]. Convolutional neural networks perform better at detecting patterns in input images such as lines, circles, gradients, and even faces and eyes. A CNN is a kind of feed-forward NN with up to twenty or thirty layers. A CNN's power is derived from a specific type of layer known as the convolution layer. CNNs are constructed from multiple convolution layers placed on top of each other and, each competent in identifying more complex structures. In this research work, performance on CNN with one, three, or four convolutional layers is observed on the dataset. A CNN's architecture is a multiple layered and feed forwarded neural network is constructed by sequentially layering multiple hidden layers on edge of one another. Convolutional neural networks can acquire hierarchical features due to their continuous architecture. Convolutional layers are usually accompanied by activation layers, and some are accompanied by pooling layers. In this research, the activation values were calculated using the ReLu function. Because the derivation of ReLu is One for affirmative input compared to typical activation functions, the ReLu function can indeed speed up deep neural network learning. With an input value a, the function can be mentioned as equation 1,

$$f(a) = returns\ maximum\ between\ 0\ and\ a \quad (1)$$

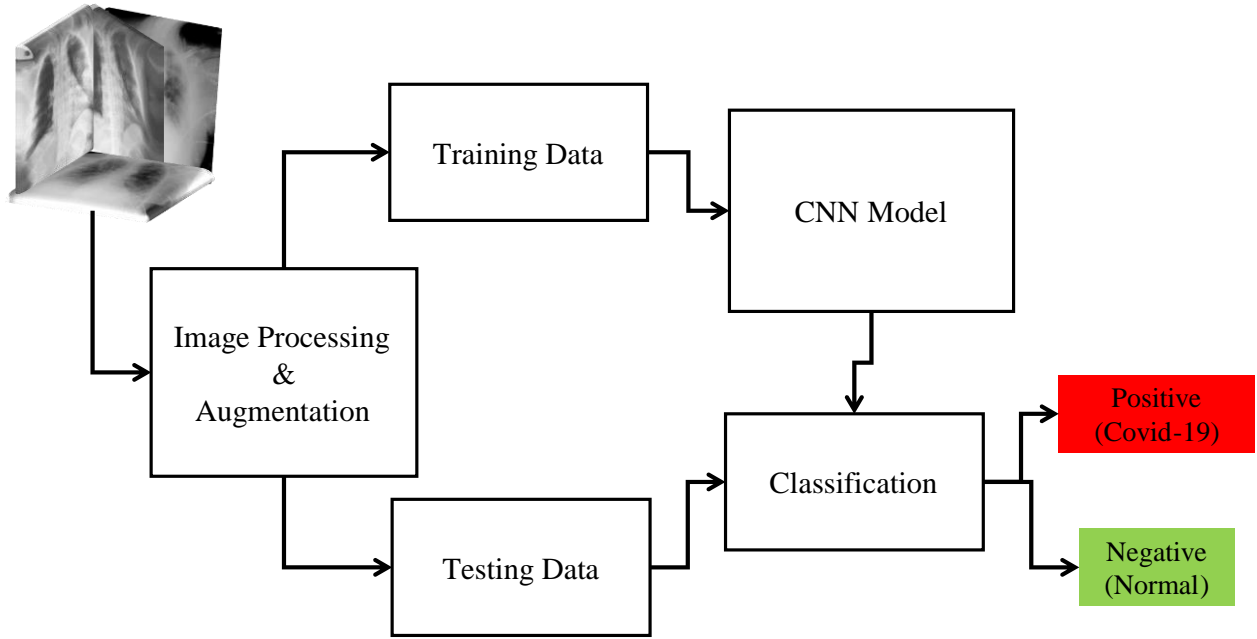

Fig. 1. System Flow Diagram of the Proposed System

The CNN is configured for this research to handle input data of dimension (30, 30, 1), which is the formatting of the dataset images. The architecture of 1, 3, and 4 layered CNN is described below.

*1) CNN with 1 Convolutional Layer:* A Conv 2D layer has been used as a 1st layer for the convolution procedure which slides a confounding filter over the input to retrieve features from the source images to generate a 3x3 feature map. For max-pooling operation, the MaxPooling2D layer of size 2 x 2 is used as the second layer. It reduces the dimensionality of each feature to shorten the time and parameters. A dropout layer is used as the third layer to combat overfitting. 20% of the neurons are disabled randomly in this research. Then dense layers are connected to feed the last output. The 3D output is flattened to 1D and connected to the classifier's process vectors. Finally, a final layer with two outputs and a softmax activation is used for 2-way classification. For this research, the loss function is binary cross-entropy, and the adam optimizer is used to update the weights of its neurons through a backpropagation. Model architecture is given in figure 2.

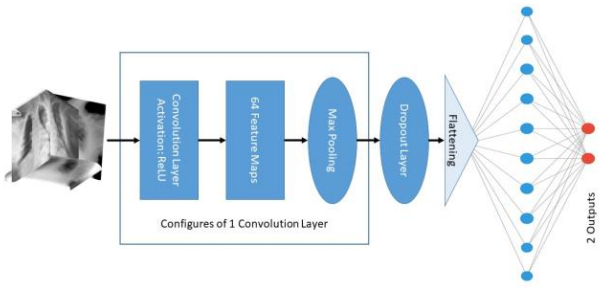

*2) CNN with 3 Convolutional Layer:* For this configuration, extra hidden layers with the first configuration of CNN are

Fig. 2. Architecture of CNN with 1 Convolutional Layer

added. There are three Conv2D layers, two Max-Pooling2D layers, and three Dropout layers are available in this configuration. 25%, 25%, 30% of the neurons are disabled randomly in this configuration for each dropout layer. Model architecture is given in figure 3.

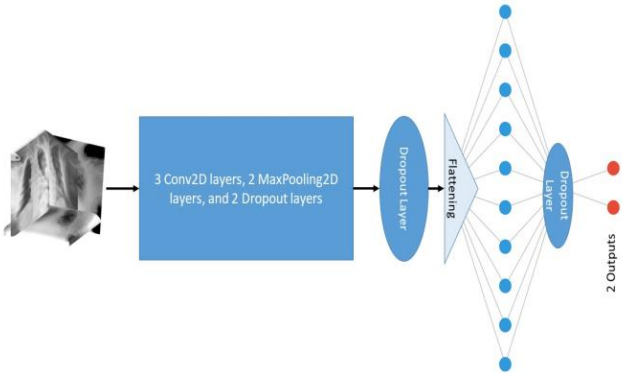

Fig. 3. Architecture of CNN with 3 Convolutional Layers

*3) CNN with 4 Convolutional Layer:* In total, this model has four Conv2D layers, two Max-Pooling layers, six batch normalization layers, and five Dropout layers with the same configuration as the first one. 25%, 25%, 25%, 40%, and 30% of the neurons are disabled randomly in this configuration for each dropout layer. Model architecture is given in figure 4.

### D. Model Training

Using the Keras deep learning library, we created a very simple Convolutional Neural Network classifier with 1, 3, and 4 convolution layers. Before being aggregated with the binary cross-entropy loss function and the Adam optimizer, the model is trained for 10 epochs with a batch size of 256. The model is then trained by an additional 50 epochs using data augmentation, which produces new training samples by rotating, zooming, and shifting on the training images.

## IV. EXPERIMENTAL RESULT

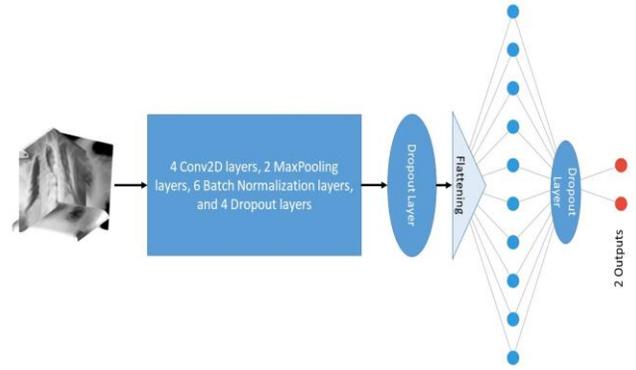

The effectiveness of a convolutional neural network with 1, 3, and 4 convolution layers for recognizing COVID-19 cases from CXR images are explored in this research. 70% of the data are used in this research study for training, 20% for the testing, and 10% for validation. Classification of normal and COVID-19 images using respective CNN models are done with and without image augmentation. Table I compares the performance of respective CNNs for a two classes classification scenario with and without image augmentation.

Fig. 4. Architecture of CNN with 4 Convolutional Layers

TABLE I
ACCURACY, PRECISION, RECALL AND F1-SCORE RESULT OF CNNs WITH AND WITHOUT IMAGE AUGMENTATION

| Image Augmentation | No of Convolution layer in CNN Model | Accuracy | Precision | Recall | F1-Score |
|---|---|---|---|---|---|
| No | 1 layer | 0.85 | 0.86 | 0.84 | 0.85 |
|  | 3 layers | 0.88 | 0.88 | 0.86 | 0.88 |
|  | 4 layers | 0.92 | 0.91 | 0.89 | 0.92 |
| Yes | 1 layer | 0.95 | 0.95 | 0.93 | 0.95 |
|  | 3 layers | 0.96 | 0.96 | 0.94 | 0.96 |
|  | 4 layers | 0.94 | 0.94 | 0.90 | 0.94 |

Based on the performance results in table I, a number of observations can be made. It is observed that with image augmentation CNN models are performing well. CNN model with three convolution layers giving the highest accuracy of 96%, where single convolution layer and four layers give the accuracy of 95% and 94% respectively. The same model achieved the highest precision, recall, and F1-score of 96%, 94%, and 96% respectively. The greater recall value obtained using the CNN with three convolution layers implies that fewer COVID-19 positive patients will be overlooked throughout the COVID screening procedure.

Figure 5 and figure 6 are showing training and validation accuracy and loss for for the best performing CNN with 3 Convolutions Layers.

In figure 5, training accuracy and validation accuracy both curves are increasing with the increase of the epochs. After epoch number around 20, training accuracy increased more than the validation accuracy.

In figure 6, training loss and validation loss both curves are decreasing with the increase of the epochs. After epoch number around 20, training loss decreased more than the validation loss. The figure 7 represents a subset of correctly predicted and figure 8 shows a subset of incorrectly predicted classes using CNN with 3 Convolutions Layers.

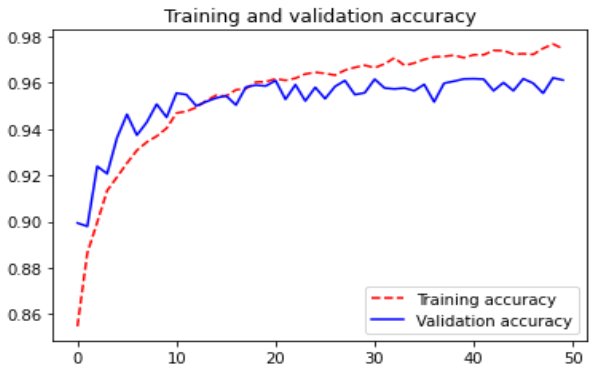

Fig. 5. Training and Validation Accuracy of CNN with 3 Convolutions Layers

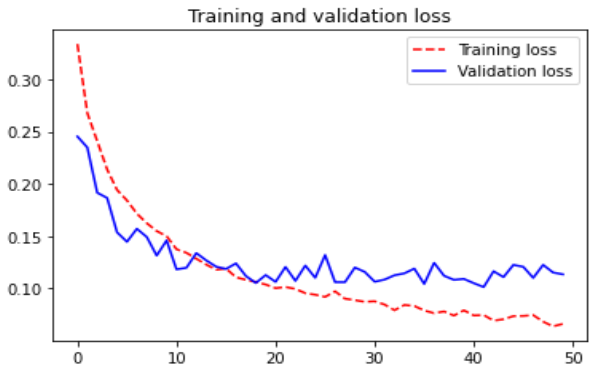

Fig. 6. Training and Validation Loss of CNN with 3 Convolutions Layers

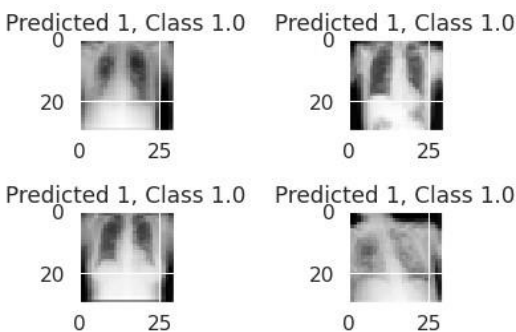

Fig. 7. Subset of Correctly Predicted Classes

## V. CONCLUSION

A 2D convolutional neural network design with three different convolution layers tailored for COVID-19 identification from CXR photographs was used in this investigation. Experiment findings show that CNN with three convolution layers might obtain significant COVID-19 detection accuracy, precision, and recall. Tweaking the model and fitting more dataset might increase the level of accuracy and precision. Through computer-aided assessment of CXR photographs of COVID19 positive patients, it has the potential to become a valuable tool for assisting doctors and front-line health professionals. If the model is trained properly from a big dataset, artificial intelligence performs magnificently in classifying COVID19. The strategy would be extremely valuable in the current pandemic, as the necessity for preventive actions conflict with existing resources.

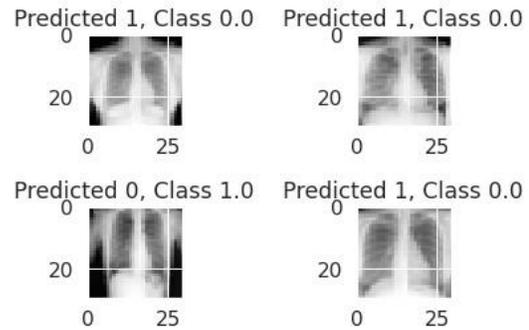

Fig. 8. Subset of incorrectly Predicted Classes